# BOOSTING STUDENTS' PERFORMANCE WITH THE AID OF SOCIAL NETWORK ANALYSIS

R.U. GOBITHAASAN*, NURUL SYAHEERA DIN, LINGESWARAN RAMACHANDRAN & ROSLAN HASNI

*School of Informatics and Applied Mathematics Universiti Malaysia Terengganu, 21030 Kuala Nerus, Terengganu*

*Corresponding author: gr@umt.edu.my

**Abstract:** There are various teaching methods developed in order to attain successful delivery of a subject without prior knowledge of the interaction among the students in a class. Social network analysis (SNA) can be used to identify individual, intermediate and group measures of interaction in a classroom. The idea is on identifying ways to boost the students' performance by means of lecturer's intervention based on their interaction. The case study was conducted involving third year batch that consisted of 76 female and 24 male students. A friendship network was drawn based on the information obtained at the end of semester 5 and it was investigated based on two metrics – centrality measures and Girvan-Newman algorithm. At the end of semester 5, grades were added as the attributes of the network. 12 clusters were found in this batch and a distinct pattern was identified between performing and poor achieving students. At the beginning of the 6th semester, the students were given the option to choose between 2 groups. Group 1 was unperturbed without any lecturer's intervention whereas the performing students' clusters in Group 1 were preserved but the students in poor performing clusters were distributed among performing clusters. The students were then asked to carry out assignments/quizzes in their respective groups. The final grades indicated that the performance of the students of Group 1 was much superior and there was clear evidence that those poor performing students in the 5th semester performed much better in semester 6. This shows that by understanding the students' interaction and incorporatinig instructor's minimal intervention, the performance of the students can be improved by creating a social contagion effect through group assignment clustering.

## Introduction

Social network is a social structure made up of nodes, ties, given weights, directions and signs. Individual, intermediate and group measures provide overall social behavior and interaction. It is defined as a group of social actors that exchange information with one another. It is a sociological approach for analyzing patterns of relationships and interactions between social actors in order to discover basic social structure (Burt & Minor, 1983). The actors can represent at various level of connectivity such as person, countries, etc. Each of the actors is represented as a node in the network and the ties between actors indicated by lines drawn between them. In a network analytic framework, the ties may be any relationship that exists between units. There are kinship, material transactions, resources flow, behavioral interaction, and the affective evaluation of one person by another. Thus, the network encompasses a knitting relationship through which the resources such as information or advise can be exchanged. However, the ties need to be established in the context of trust and friendship that may help to achieve a particular objective among the nodes.

A number of software applications have been developed to carry out social networks analysis (SNA). Commonly used applications are Pajek (Batagelj and Mrvar, 2006), UCINET (Borgatti et al., 1999) and R (R Development Core Team, 2007). Recenty, SNA has become a subject grouped under data analytics program which involves crunching big data for effective decision making.

Friendship is a fundamental characteristic of human relationships whereas individuality is generally presumed to be reciprocal in nature. Students form friendship ties by nature that influence them in many ways; friends can motivate and support but also can be highly disruptive. It is a type of a network which has social influence and these kinds of networks have the capability to recreate, for example, the diffusion of new behaviors, new ideas and new products in a network (Almaatouq et. al, 2016).

In this research, we utilized UCINET to draw and analyze the sociology of a batch of Computational Mathematics students in 5th/6th semester and further elucidate their interaction that led to assignment group formation and final grades. The objectives were twofold; firstly, we wished to identify the existing friendship clusters (5th semester) of Computational Mathematics students and we further added two attributes which were grade and gender to the network. Secondly, we investigated the changes of grade patterns by perturbing the existing friendship clusters; students' assignment subgroups were perturbed in the 6th semester where the existing performing students' assignment subgroups were preserved but the students in poor achieving assignment subgroups were distributed among performing assignment subgroups. The students were then asked to carry out assignments/quizzes in their new respective groups and the final grades were investigated.

**5th Semesters' Friendship Network**

Figure 1 represents a friendship network constructed using a set of data collected from 100 students of Computational Mathematics of Universiti Malaysia Terengganu at the end of semester 5 (2017/2018). When the data was collected, they had already spent two years together and hence all of them knew each other. Each student was asked who they spent time with. They were provided with the list of students in the class. Since the question was an asymmetric relation, the final data produced a directed graph where each node (V) represented a student and the ties (E) showed the interconnection between them. This is due to the fact that not all friendships are reciprocal (Varquera et al., 2008). The network was drawn with UCINET using the adjacency matrix produced from the data.

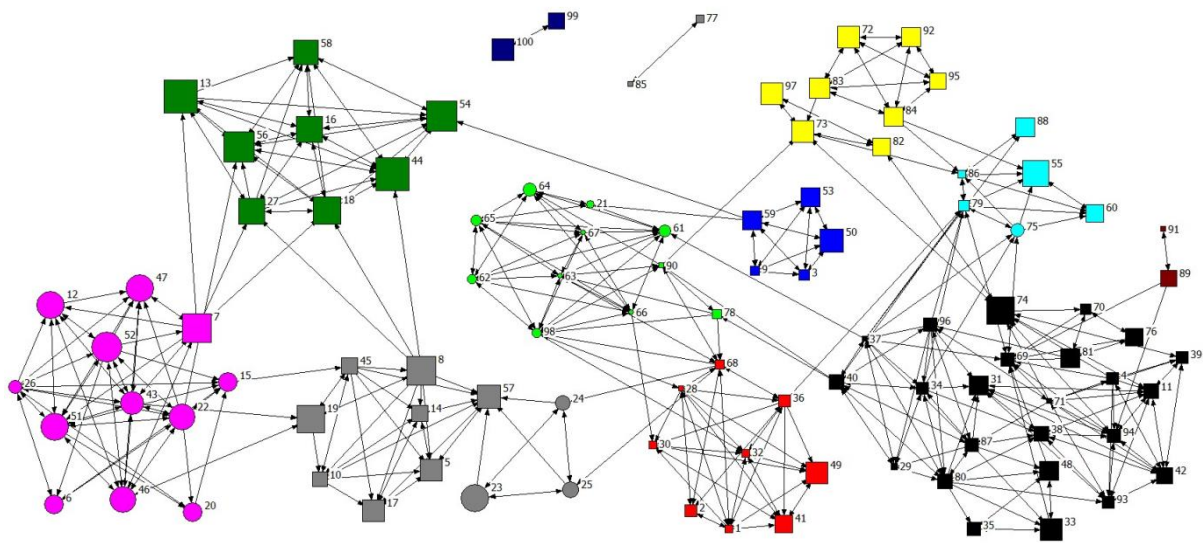

Figure 1: Friendship network using UCINET, denoted as F5(V,E) for Geometric Modelling taught in semester 5.

The nodes were number coded to replace students' names to protect their privacy. There were 100 students with 511 ties. The first attribute added to the graph was the gender in the form of the shape of the node. There were 24 male students represented using circular nodes) and 76 female students represented with square nodes. The second attribute was the size of the node; it varied based on their final grade of the course entitled Geometric Modelling where the bigger the node size, the higher the grade obtained. The average mark was 62% where only 50 students managed to score more 70% or higher. Meanwhile, the university has set out a new academic goal of achieving 70% average for each course taught in UMT; thus this batch did not meet the goal.

It is evident from Figure 1 that there were three distinct network components in this batch. There were two networks with only a pair of students and the grey pair was a low achieving one. The male cluster was colored with light green which was a part of the main component of the network that was also considered as a low achieving group. It was the same for the red female cluster where the average grade of this cluster was equally low. The brown cluster consisted of a pendant vertex for the main component; these were students with comparatively poor grade. The two tiny components with a pair of students could not be categorized as antisocial for not being directly connected to the main component as the students had friends from other undergraduate programmes and it was similar for the students denoted as the pendant vertex. The higher performing groups were the green and pink clusters.

Figure 2 shows a skewed distribution of students' grades which indicated the average grade did not even reflect the total grades. It also indicates that there was a clear disparity between high performing students and low achieving students. This indicates that those interested in the subject matter worked hard with minimal attention given to them. Nonetheless, those without any interest in the subject were likely to score lower grade regardless of extra effort given by the instructor. However, this was a mere postulate derived from 9 years of teaching experience at the university which needs further clarification.

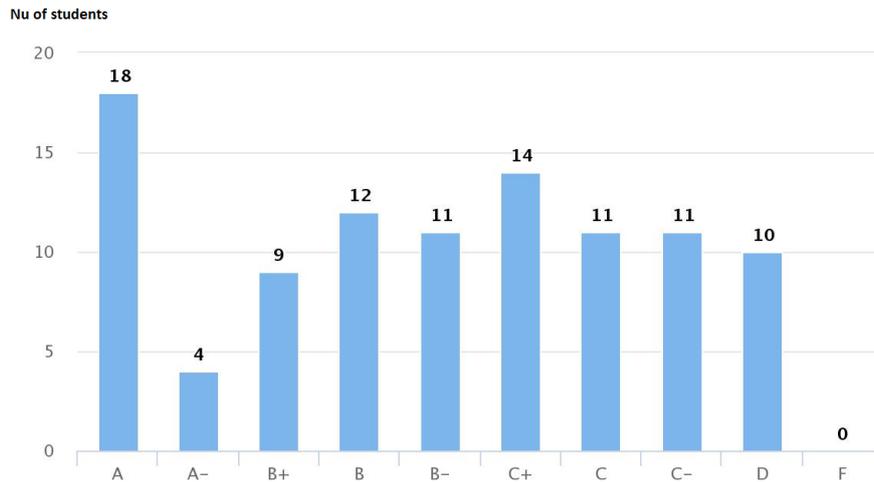

Figure 2: Grade distribution for Geometric Modeling of F5(V,E).

The nodes in Figure 1 were color coded; it was the third attribute where the students were grouped based on Girvan-Newman partition measure with the highest score of having 12 clusters, denoted as $Q_{12} = 0.720$. The following lists denotes clusters' number with subscript and its partition measure:

{ $Q_3 = 0.012$, 4 clusters: $Q_4 = 0.468$, $Q_5 = 0.629$, $Q_6 = 0.649$, $Q_7 = 0.669$, $Q_8 = 0.700$, $Q_9 = 0.698$, $Q_{10} = 0.714$, $Q_{11} = 0.717$, $Q_{12} = 0.720$, $Q_{13} = 0.713$, $Q_{14} = 0.714$, $Q_{15} = 0.712$ }.

It was fairly easy to appoint a class representative by computing centrality measures of the network. As it was a disconnected network with three components, closeness centrality could not be employed. It was the same for eigenvector centrality measure where the centrality measure could be misleading for directed network (Borgatti et al., 2002). Figure 3 shows an example of identifying class representative using directed betweenness centrality where the bigger the size, the higher its betweenness measure. Node 74 had the highest betweenness value of 842.901, followed by node 71 (774.898) and node 73 (742.544). These three nodes were suitable to be pointed as a class representative; they were just a few nodes away from other students in the main component. Even though node 90 (678.577) seemed to be in the center to reach all the nodes, in actuality it ws ranked fifth after node 94 (688.288).

Based on the formation of friendship network, it is crucial for instructors to carefully carve a plan to intervene and create a peer support system which enables knowledge sharing among students. Numerous domains have employed this system influence and support behavioral changes, for example smoking/alcohol cessation, weight loss, diabetes management etc. (Almaatouq et. al, 2016).

A notable fact of UMTs' grading marking scheme is 60% is contributed by marks from coursework assignments, quizzes, laboratory/tutorial exercises and from midterm tests. The remaining 40% is contributed by marks from final examinations. Hence, students' grades are likely to increase if they are carefully grouped, with the hope that intelligent students may help poor performing students.

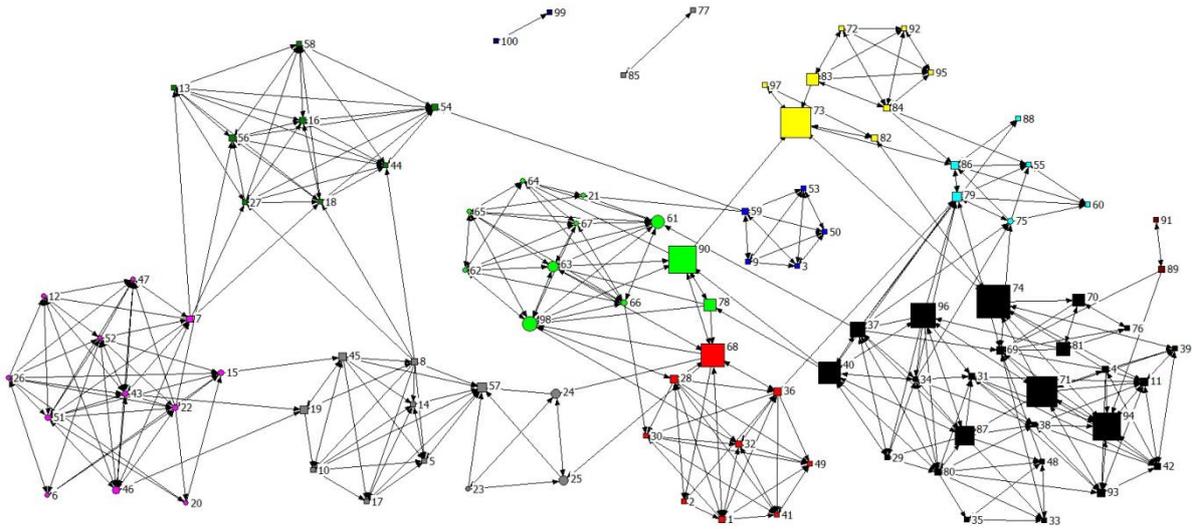

Figure 3: Identifying class representative for F5(V,E) using directed betweenness centrality.

**Clustering Students in 6th Semesters for Group Assignment**

In semester 6, the same batch consisting of 100 students were taught a new course called Advanced Numerical Analysis. As the faculty administration decided to break big lecture groups into smaller ones, this batch was divided into two groups. The students were given the option to either choose Group 1 or Group 2. The first author of this paper taught Group 1 consisting of 56 students whereas 44 students chose group 2 which was taught by a new instructor with similar materials. Group assignment members for Group 1 were predetermined by the instructor based on the friendship network F5(V,E) whereas Group 2 was the control group without instructors' intervention. High performing groups were unperturbed; we maintained their togetherness in the 6th semester and dissolved low achieving groups by dispersing them to high achieving performing groups.

Figure 4 illustrates friendship network of Group 1 denoted as F6a(V, E) consisting of 16 male and 40 female students. There were 5 types of nodes' shapes indicating five new assignment groups. However, the colors of nodes were chosen based on the clusters' colors as shown in F5(V,E). The friendship ties were derived from 5th semester as well. The formation of Group 1 was consistent with the number of reciprocal ties in a cluster where reciprocal ties tended to stick together in the 6th semester by nature, especially the high performing groups where the majority of the ties were in the form of reciprocal. The clusters with less reciprocal ties tended to break during the division. Most of the black, yellow, grey and cyan clusters chose Group 2. The size of the nodes indicated the marks they obtained in the 5th semester for Geometric Modeling where the average mark for Group 1 was 65.4%.

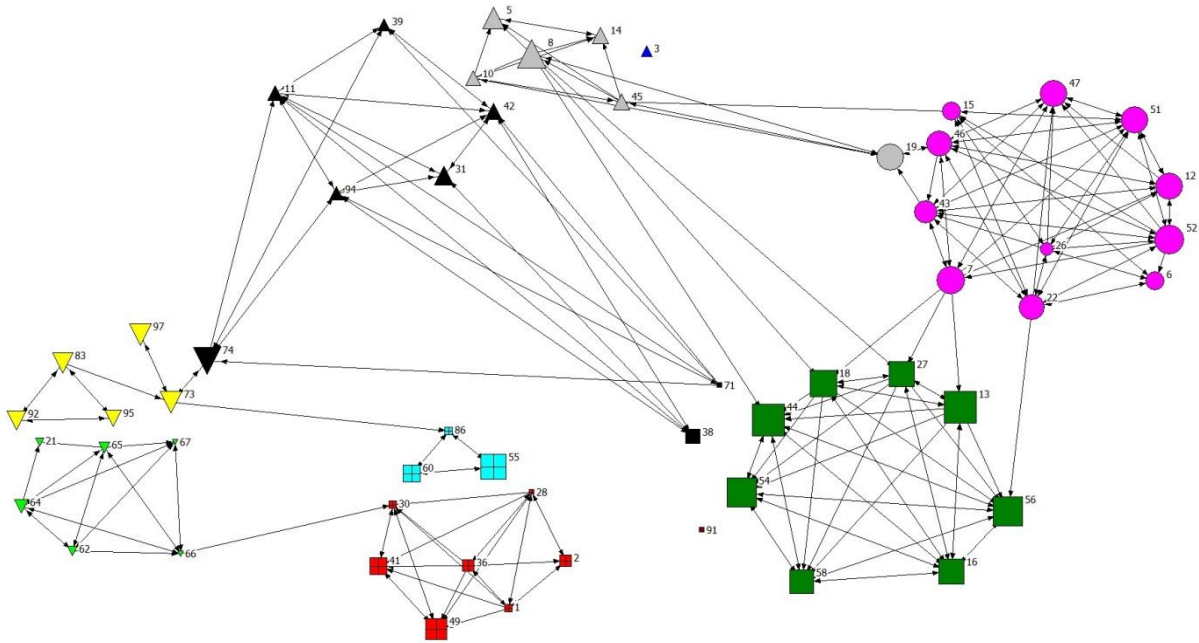

Figure 4: F6a(V, E) shows Group 1 students with their existing friendship ties and marks.

Figure 5 denoted as F6b(V,E) shows the same network of F6a(V,E) but with 6$^{th}$ semesters' Advanced Numerical Analysis marks representing the size of nodes. The average score of this group increased to 70.8%. Those poor performing nodes especially node 91, 3 and 71 showed improvement in their final score. Similarly, the boys' cluster (light green) showed drastic improvement except for node 66. However, the average performing students in black and grey clusters showed no improvement as well with the lowest marks obtained by node 39 (46%).

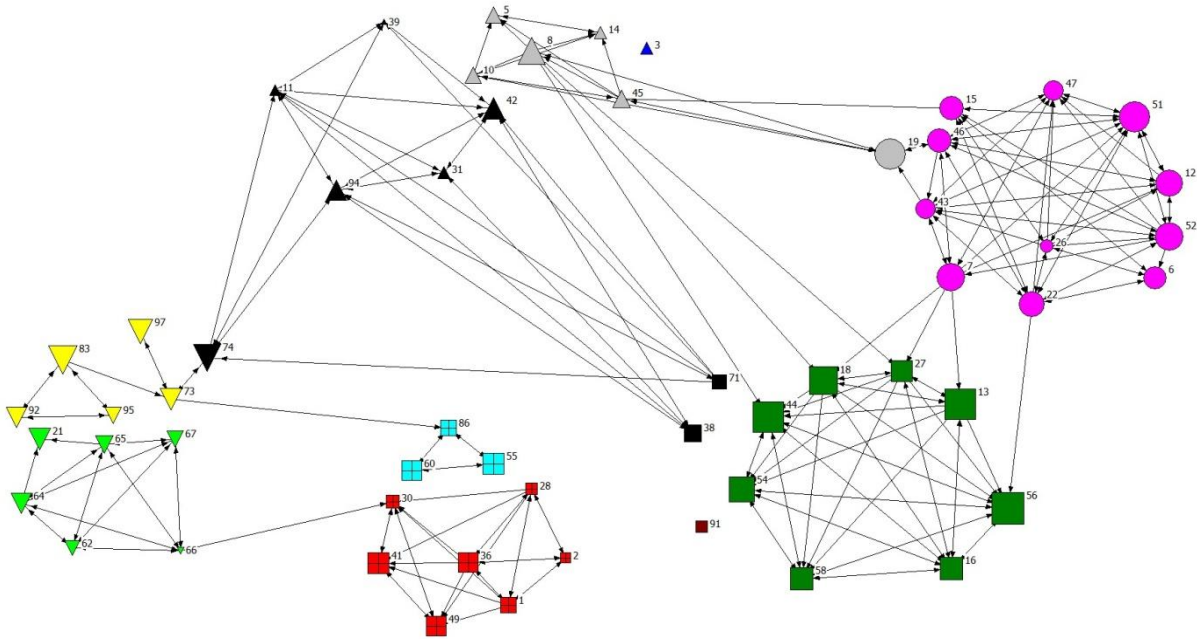

Figure 5: Groups 1 with its original friendship network denoted as F6b(V, E)

Figure 6 below shows the grade distribution of Group 1 and Figure 7 shows the grade distribution for the control group. The grade for Group 1 had a skewed distribution indicating the average score did not reflect the grades, whereas Group 2 had a normal distribution indicating the average score was close to the mean and mode.

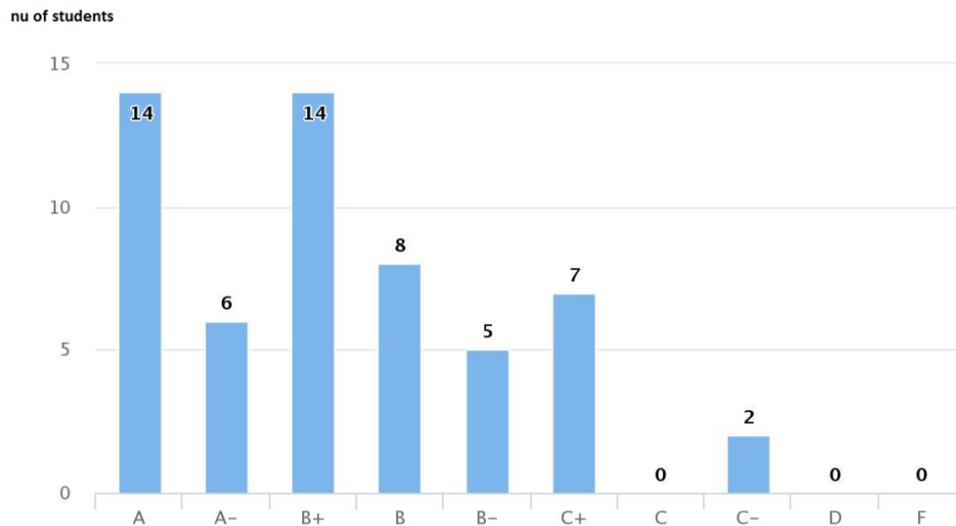

Figure 6: Grade distribution for Group 1 as shown in F6b(V,E)

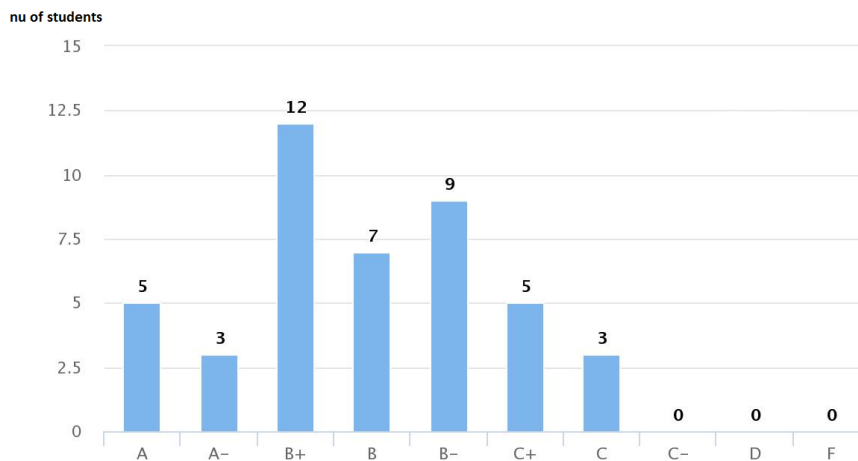

Figure 7 Grade distribution for Group 2

**Conclusion**

In this study we demonstrated how the measures available in SNA, especially centrality and Girvan-Newman clustering measures could be used to create a social contagion effect, where lower achieving students were grouped with high performers to increase the average marks of the poor performing students. A positive influence was assured if the cluster consisted of numerous high performers. However, no improvement could be found if the poor achieving student was attached with average performing clusters where the exertion to study failed. In the case of the boys' cluster, the improvement was further boosted if subgroups of low performing students were kept undivided before attaching them to a high performing cluster. Regardless of these changes, a number of poor achieving students maintained low marks indicating there were no significant behavioral changes. However, the final grade was a right skewed distribution, thus the average score could not be used to reflect their grades. Future work includes identifying how the strength of friendships (reciprocity) can impact the influence on the individuals or clusters.

**Acknowledgements**

The authors acknowledge the Ministry of Education for providing FRGS (59431) grant and they further acknowledge Dr. Maharani for providing marks for Group 2 which was utilized as the control group.